\documentclass[%
 reprint,
 amsmath,amssymb,
 aps,
]{revtex4-2}
\pdfoutput=1
\usepackage{dcolumn}
\usepackage{bm}
\usepackage{amsfonts}
\usepackage{amssymb}
\usepackage{amsmath}    
\usepackage{graphicx}   
\usepackage{verbatim}   
\usepackage{color}      
\usepackage[colorlinks=true,linkcolor=blue,citecolor=blue,allcolors=blue]{hyperref}
\usepackage{lineno}
\usepackage{lipsum}
\usepackage{multirow}
\usepackage{afterpage}
\usepackage{flushend}
\usepackage[export]{adjustbox}
\usepackage{float}
\usepackage{footmisc}
\usepackage[english]{babel}
\usepackage[utf8]{inputenc}
\usepackage[section]{placeins}
\usepackage{xr}
\usepackage{pifont}
\makeatletter
\newcommand*{\rom}[1]{\expandafter\@slowromancap\romannumeral #1@}
\makeatother

\begin{document}



\title{Droplets in Acoustic Fields: A Unified Theory from Migration to Splitting}
\author{ Jeyapradhap Thirisangu,$^{1\dagger}$ \ Varun Kumar Rajendran,$^{1\dagger }$\  Snekan Selvakannan,$^1$\  Sujith Jayakumar,$^1$\ E Hemachandran,$^2$  \ and Karthick Subramani$^1$} 
\email{karthick@iiitdm.ac.in}
\affiliation{$^1$ Department of Mechanical Engineering, Indian Institute of Information Technology, Design and Manufacturing Kancheepuram, Chennai-600127, India.} 
\affiliation{$^2$ Department of Foundry and Forge Technology, National Institute of Advanced Manufacturing Technology, Ranchi – 834 003, Jharkhand, India.} 
\begin{abstract}
We present a comprehensive theoretical framework governing the dynamics of droplets in acoustic fields, applicable to all droplet sizes, from the Rayleigh limit ($D<<\lambda$) and beyond ($D\gtrsim  \lambda$). Our theory elucidates the nature of acoustic forces and incorporates the effects of interfacial tension, enabling predictions of droplet migration, deformation, and splitting under suitable conditions. Importantly, we demonstrate droplet deformation and splitting through bulk acoustic wave (BAW) silicon-glass microscale experiments, which validate the proposed theory.

\end{abstract}

\maketitle




 Controlling and manipulating droplets using acoustic fields has applications in various fields, including microfluidics \cite{Friend2011Jun}, biotechnology \cite{Wu2019Jun, Rufo2022Apr,Rufo2022Jun}, material science \cite{Foresti2018Aug}, and chemical engineering \cite{Li2019Jan}. Despite the remarkable progress in experiments \cite{Foresti2013Jul,Tian2016Oct,Leibacher2015Jun} and applications \cite{Friend2011Jun,Wu2019Jun, Rufo2022Apr,Rufo2022Jun,Foresti2018Aug,Li2019Jan,Zhang2018Jul}, the theoretical understanding and analysis of droplets subjected to the acoustic field are still largely restricted to Gor'kov theory of acoustic radiation force ($\boldsymbol{F}_{ac}$)  acting on particles, which considers droplets as small rigid particles \citep{king1934acoustic, gor1962forces, yosioka1955acoustic,Crum1971Jul,Liu2022Apr}. 
Employing $\boldsymbol{F}_{ac}$ to study the droplets has key limitations as follows: 1) it is valid only if the droplet size ($D$) is much smaller than the wavelength ($\lambda$) of the wave $(D<<\lambda)$. 2) the $\boldsymbol{F}_{ac}$ is in the Lagrangian form, which focuses on individual particle behavior and is not directly compatible with the Navier-Stokes equations to study the interfacial effects. As a result, the  $\boldsymbol{F}_{ac}$ can only predict migration of small rigid droplets and cannot be employed to theoretically investigate phenomena that involves interfacial effects such as droplet deformation, breakup and coalescense under acoustic fields. Recently, there has been an emphasis on the theoretical understanding of acoustic radiation forces acting on large droplets \cite{Baasch2020Jan,PazosOspina2022Sep, Hasegawa1969Nov,Marston2017Sep}. However, these studies also neglected the interfacial effects by treating droplets as particles and primarily focused on determining the spatial averaged acoustic radiation force in Lagrangian.

Moreover, the studies on droplet deformation and breakup under acoustic levitation \cite{Lee1991Nov, Lee1994Nov, Yarin1998Feb,Tian1993Jun,Shi1996Apr,Di2018Oct,Naka2020Dec,Zang2017Mar} highlight the shortcomings of the current theoretical framework employed to study the droplets in acoustic fields. In these studies, the acoustic force ($\boldsymbol{F}_{ac}$) is employed to calculate the migration and positioning of levitated droplets but not their deformation. Droplet deformation is typically determined by substituting acoustic radiation pressure on the droplet surface into the Young-Laplace equation. This approach, however, only provides the static deformation shape and disregards the dynamics. Furthermore, the theoretical framework \cite{Stone1989Jan, Danilov1992Nov,Foresti2013Jul,Naka2020Dec, Aoki2020May} to study droplet splitting under acoustic fields is still nonexistent.

The main goal of this work is to establish the unified theoretical framework with Eulerian form of acoustic force to study and explore the behaviour of droplets under acoustic fields. Remarkably, we demonstrate the deformation, migration, and splitting of droplets subjected to standing acoustic waves through silicon-glass microchannel experiments, thereby confirming the theoretical predictions.


The dynamics of inhomogeneous immiscible fluids subjected to acoustic waves considered in this study is governed by mass-continuity, momentum, and advection equations and thermodynamic pressure-density relation \citep{Landau1987Aug}, which are given below,
\begin{subequations}
\label{Eq 1}
\begin{equation}
\label{Eq 1a}
    \partial_t \rho+\boldsymbol{\nabla} \cdot\left(\rho \textbf{\emph{v}}\right)=0,
\end{equation} 
\begin{equation}
\begin{split}
\label{Eq 1b}
    \rho[\partial_t \textbf{\emph{v}} + (\textbf{\emph{v}}\cdot\boldsymbol{\nabla})\textbf{\emph{v}}]=-\boldsymbol{\nabla} p +\eta\boldsymbol{\nabla}^2 \textbf{\emph{v}}  +\beta\eta\boldsymbol{\nabla}(\boldsymbol{\nabla}\cdot\textbf{\emph{v}}) 
    \\
    +\rho\textbf{\emph{g}}+\textbf{\emph{f}}_{\sigma},
\end{split}
\end{equation}
\begin{equation}
\label{Eq 1c}
    \partial_t \phi+\textbf{\emph{v}}\cdot\boldsymbol{\nabla} \phi = 0,
\end{equation}
\begin{equation}
\label{Eq 1d}
   \frac{d\rho}{d t}=\frac{1}{c^2}\frac{dp}{d t},
\end{equation}
\end{subequations}

where $\rho$ is the density, $\textbf{\emph{v}}$ is the velocity, $p$ is the pressure, $\eta$ is the dynamic viscosity of the fluid, $\xi$ is the bulk viscosity, $\beta=(\xi/\eta)+(1/3)$, ${\phi}$ is the phase fraction ($\phi=0$ for fluid $1$ and $\phi=1$ for fluid $2$), 
$\rho\textbf{\emph{g}}$ is the gravity force and $c$ is the adiabatic local speed of sound ($c^2=(\partial p/\partial\rho)|_{S}$). The interfacial force ($\textbf{\emph{f}}_{\sigma}$) is expressed as $\sigma \kappa \delta_{s} \textbf{n}$, where $\sigma$ is the surface tension, $\kappa$ is the curvature, $\textbf{n}$ is the unit normal and $\delta_{s}$ is a surface Dirac delta function which is non-zero only on the interface.
The applied high-frequency acoustic fields, particularly in the MHz range and above, vary on fast time scales ($t_f \sim 1/\omega \sim$ 0.1 \textmu s). These fast time acoustic fields can induce slow time scale, $t_s$ ($t_s\gg t_f$) phenomena such as acoustic radiation force, streaming, and relocation of inhomogeneous fluids. Therefore, all the fields can be analyzed by decomposing them into slow time fields ($g_s$) and fast time fields (${g}_{f} e^{-i\omega t_f}$), as follows \cite{Eckart1948Jan}: 
\begin{equation}
\label{Eq 2}    g=g_s(\textbf{\emph{r}},t_s)+g_f(\textbf{\emph{r}},t_s)e^{-i\omega t_f}.
\end{equation}
In a homogeneous fluid, the amplitude of the fast time fields ($g_f$) remains constant and does not change with the slow time scale ($t_s$). In contrast, in an inhomogeneous fluid,  $g_f$ becomes dependent on the slow time scale. This dependence arises because $g_f$ is affected by spatial variations in density ($\rho_s$) and speed of sound (${c}_{s}$), which themselves vary over the slow time scale. The rate of change of slow time scale fields is much smaller than the rate of change of fast time fields ($\frac{\partial g_s}{\partial t}\ll\frac{\partial g_f}{\partial t}$). To derive the fast and slow time governing equations, the following assumptions are employed:
1) The magnitude of the fast time scale velocity field is much smaller than the speed of the sound of the medium ($\left|\textbf{\emph{v}}_f\right|\ll\ {c}$) or acoustic Mach number, $ma_f = \left|\textbf{\emph{v}}_f\right|/{c}\ll\ 1$.
2) The slow time scale velocity magnitude is much smaller than the fast time scale velocity, $\left|\textbf{\emph{v}}_s\right| \ll\ \left|\textbf{\emph{v}}_f\right|$ or the hydrodynamic Mach number, $ma_s = \left|\textbf{\emph{v}}_s\right|/{c}\ll\ ma_f$.
3) The gravity and interfacial forces can only contribute to the slow time scale fields and create negligible effects on fast time scale fields.
In this work, the assumption $\boldsymbol{\nabla} p_s \ll \boldsymbol{\nabla} p_f$ in fast time scale equations cannot be employed, as in previous studies \cite{Rajendran2022Jun}, due to a sharp pressure jump $p_s$ across the interface induced by interfacial forces.
Substituting Eq. \ref{Eq 2} in Eqs. \ref{Eq 1}, along with the above assumptions, yields the following governing equation of the fast time fields,
\begin{subequations}
\label{Eq 3}
\begin{equation}
   \tag{3a}\label{Eq 3a}
    \partial_t \rho_f +\boldsymbol{\nabla} \cdot(\rho_s \textbf{\emph{{v}}}_f)=0,
\end{equation}
\begin{multline}
\tag{3b}\label{Eq 3b}
    \rho_s \partial_t \textbf{\emph{v}}_f
    =-\boldsymbol{\nabla} (p_s+p_f) + \eta\boldsymbol{\nabla}^2 \textbf{\emph{v}}_f+\beta\eta \boldsymbol{\nabla}(\boldsymbol{\nabla}\cdot\textbf{\emph{v}}_f),
\end{multline}
\begin{equation}
\tag{3c}\label{Eq 3c}
    \partial_t\phi_f+\textbf{\emph{v}}_f\cdot\boldsymbol{\nabla} \phi_s =0,
\end{equation}
\begin{equation}
\tag{3d}\label{Eq 3d}
    \partial_t\rho_f+(\textbf{\emph{v}}_f\cdot\boldsymbol{\nabla})\rho_s=(1/c^2)\partial_t p_f.
\end{equation}
\end{subequations}

\begin{figure*}[t!]
  \center    \includegraphics[width=0.9\linewidth]{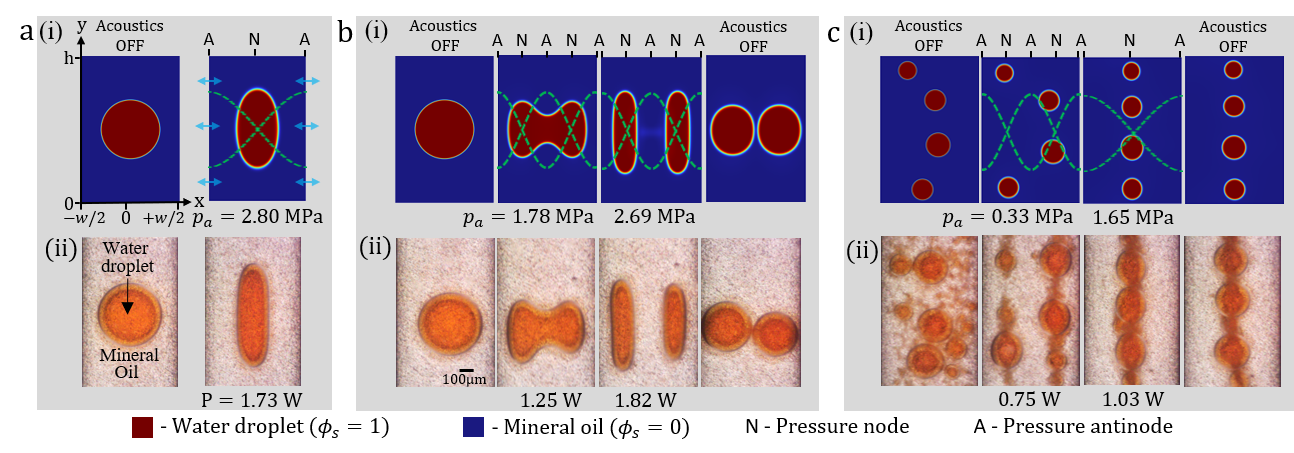} 
    \caption{ Droplets behaviour in a standing acoustic wave. (a) Droplet deformation in a half-wave ($\lambda \simeq  2w$): (i) Theoretical predictions at $f = 1.046$ MHz and $a = 1$ nm, and (ii) experiments at $f =0.987 $ MHz and power applied to the transducer (P) $= 1.73$ W. (b) Droplet deformation and splitting in a full-wave ($\lambda \simeq  w$): (i) Theoretical predictions, at
    $f=2.03$ MHz and $a=13.9$ nm yields droplet deformation towards both nodes ($p_a=1.78$ MPa) and at $f=2.03$ MHz and $a=14$ nm, droplet splits ($p_a=2.69$ MPa) resulting in two daughter droplets. (ii) experiments at $f = 1.954$ MHz and P $= 1.82$ W yields  droplet splitting. 
    (c) Migration of smaller droplets to nodes with negligible deformation: (i) Theoretical predictions of droplets exposed to a full-wave ($f = 2.05$ MHz and $a = 2.1$ nm) then a half-wave ($f = 1.05$ MHz and $a = 2.1$ nm), and (ii) experiments with droplet suspensions exposed to a full-wave ($f =1.954$ MHz and P $= 0.75$ W) then a half-wave ($f =0.987 $ MHz and P $= 1.03$ W). It is important to note that all the experiments were conducted using only a $1$ mm thick hard PZT transducer and a $670$ µm microchannel, with frequency and power as the only variables.}   
\label{Fig 1}
\end{figure*}
\par The fast time scale fields vary harmonically with time, leading to a time-averaged value of zero. Consequently, the fast time scale terms in Eqs. \ref{Eq 3} do not induce any bulk fluid motion. To obtain the governing equation for the slow time scale flows, substitute Eq. \ref{Eq 2} in Eqs. \ref{Eq 1}, apply the time averaging of first-order fields over the one oscillation period or up to $t_s \gg t_f$ and combining the continuity equation and pressure-density equation leads to the following slow time equations,
\begin{subequations}
\label{Eq 4}
\begin{equation}
   \tag{4a}\label{Eq 4a}
    \boldsymbol{\nabla}\cdot \textbf{\emph{v}}_s =0,
\end{equation}
\begin{multline} 
\tag{4b}\label{Eq 4b}
   \rho_s \partial_t \textbf{\emph{v}}_s  + 
 \rho_s \textbf{\emph{v}}_s \cdot\boldsymbol{\nabla} \textbf{\emph{v}}_s = -\boldsymbol{\nabla} p_s + \eta\boldsymbol{\nabla}^2 \textbf{\emph{{v}}}_s  - \langle \rho_f \partial_t \textbf{\emph{v}}_f \rangle \\- \langle \rho_s \textbf{\emph{v}}_f \cdot\boldsymbol{\nabla} \textbf{\emph{v}}_f \rangle+\rho_s \textbf{\emph{g}} +\sigma k\delta_{s} \textbf{n},
\end{multline}
\begin{equation}
\tag{4c}\label{Eq 4c}    \partial_t\phi_s+\textbf{\emph{v}}_s\cdot\boldsymbol{\nabla} \phi_s =0,
\end{equation}
\end{subequations}
where $\langle ... \rangle$ represents time-average over one oscillation period. For detailed derivation of Eqs. \ref{Eq 3} and \ref{Eq 4} refer to supplementary material. Slow time Eqs. \ref{Eq 4} and fast time Eqs. \ref{Eq 3} together govern the dynamics of droplets (inhomogeneous immiscible fluids) in acoustic fields. For droplets suspended in a microchannel under standing waves, the above equations predict deformation, migration, and splitting of the droplets under various conditions as shown in Fig. \ref{Fig 1}. Remarkably, our experimental observations validate all these theoretical predictions (Fig. \ref{Fig 1}). It must be noted that the effect of gravity is neglected in this microscale study, and it should be accounted for in levitation studies. The governing equations Eqs. (\ref{Eq 3} and \ref{Eq 4}) are bidirectionally coupled and solved in COMSOL Multiphysics 6.0 to obtain theoretical results (Fig. \ref{Fig 1}). An acoustic standing wave along the x-direction is imposed in the microchannl by actuating both side walls at a specific frequency $'f'$ and wall displacement $'a'$ in the x-direction. In simulations, the interface is modeled as diffusive to avoid singularities and enhance stability by incorporating a diffusion term in Eq. \ref{Eq 4c}. For detailed simulation and experimental procedures, refer to supplementary material.

Before delving into a thorough analysis of the interplay between acoustic ($\textbf{\emph{f}}_{ac}$) and interfacial forces ($\textbf{\emph{f}}_{\sigma}$), these findings can be roughly explained as follows: Acoustic forces try to move high-impedance droplet/fluid to the nearest node. If $\textbf{\emph{f}}_{ac} << \textbf{\emph{f}}_{\sigma}$, the droplet migrates to the node and behaves like a rigid particle without noticeable deformation (Fig. \ref{Fig 1}c). In contrast,  significant droplet deformation occurs if $\textbf{\emph{f}}_{ac} \gtrsim \textbf{\emph{f}}_{\sigma}$. For droplets smaller than $\lambda / 2$, high-impedance droplets are squeezed towards the node (Fig. \ref{Fig 1}a). If the size of a droplet exceeds $\lambda / 2$, multiple nodes can exist within the droplet, then $\textbf{\emph{f}}_{ac}$ act to squeeze the droplet toward each node. If the $\textbf{\emph{f}}_{ac}$ is sufficiently strong, this deformation can potentially lead to the droplet splitting (Fig. \ref{Fig 1}b). 

\begin{figure}[h!]
  \center
    \includegraphics[width=1\linewidth]{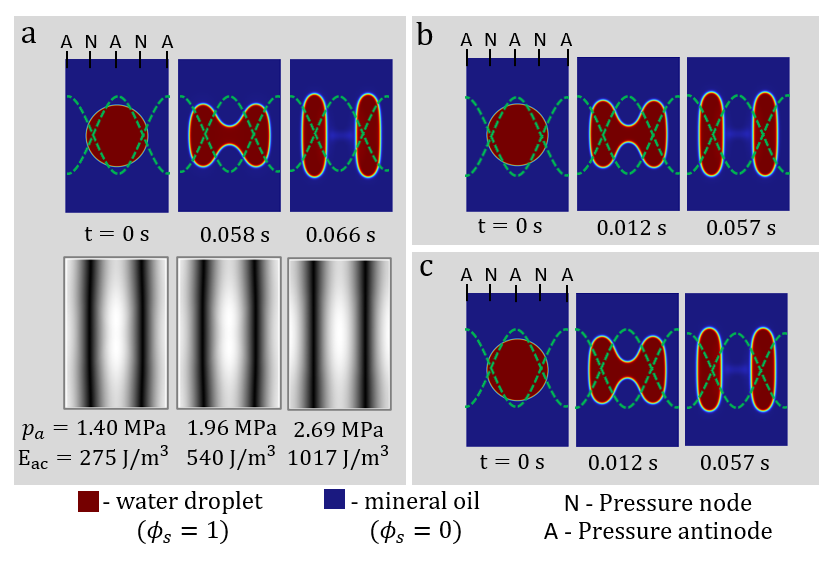} 
    \caption{Results of droplet splitting using a) standard theoretical framework that solves both fast acoustic (Eqs. \ref{Eq 3}) and slow hydrodynamic fields (Eqs. \ref{Eq 4}), at $f=2.03$ MHz and $a=14$ nm. The greyscale image shows a fast time pressure acoustic field (black: node and white: antinode) that changes with slow time. b) and c)  simplified theoretical framework using Eq. \ref{Eq 8} ($\textbf{\emph{f}}_{ac1}$) and \ref{Eq 12} ($\textbf{\emph{f}}_{ac2}$) respectively, at a steady $E_{ac}=$ 540 J/m$^{3}$.  Simplified frameworks solve only slow time fields.}
\label{Fig 3}
\end{figure}

A thorough analysis follows to elucidate the results comprehensively. The time-averaged terms in momentum Eq. \ref{Eq 4b} (Eulerian form) acts as an acoustic body force ($\textbf{\emph{f}}_{ac}$), can also be expressed as the divergence of the Reynolds stress tensor,   
    \begin{equation}
    \begin{split}
 \label{Eq 5}
   \textbf{\emph{f}}_{ac}& =-\langle \rho_f \partial_t \textbf{\emph{v}}_f \rangle 
    - \langle \rho_s \textbf{\emph{v}}_{f} \cdot\boldsymbol{\nabla} \textbf{\emph{v}}_f \rangle = -\boldsymbol{\nabla}\cdot\langle\rho_s\textbf{\emph{v}}_f \otimes \textbf{\emph{v}}_f\rangle.
\end{split}
\end{equation} 
The above equation can be simplified further by neglecting the acoustic streaming effects as follows \cite{Rajendran2022Jun}, 
\begin{equation}
\begin{split}
\label{Eq 6}   
    \textbf{\emph{f}}_{ac}& = \frac{1}{2}\boldsymbol{\nabla}(\kappa_s\langle|p_f|^2\rangle-\rho_s \langle|\textbf{\emph{v}}_f|^2\rangle) \\&
    -\frac{1}{2}(\langle|p_f|^2\rangle\boldsymbol{\nabla}\kappa_s+\langle|\textbf{\emph{v}}_f|^2\rangle\boldsymbol{\nabla}\rho_s)
    =\boldsymbol{\nabla}\phi_{{ac}}+
\textbf{\emph{f}}_{{ac1}}.
\end{split}
\end{equation}
For the standing acoustic wave applied along the x-direction, (pressure $p_f=p_a\sin(kx)$ and velocity $\textbf{\emph{v}}_f=\frac{p_a}{i\rho_s c_s}\cos(kx) $) the non-gradient body force term $({f}_{{ac1}})$ becomes \cite{Rajendran2022Jun}, 
\begin{equation}
\label{Eq 7}
   \textbf{\emph{f}}_{ac1}=-E_{ac}\cos({2kx})\boldsymbol{\nabla}\hat{Z},
\end{equation}
where $E_{ac}=p_a^2/(4\rho_{avg}c_{avg}^2)$ is the acoustic energy density, $p_a$ is the pressure amplitude, $Z = \rho_s c_s$ denotes impedance, $\hat{Z}=Z/Z_{avg}$. By substituting Eq. \ref{Eq 6} and Eq. \ref{Eq 7} into the Eq. \ref{Eq 4b} and rewriting surface Dirac delta function as Heaviside step function ($H$) results in, 
\begin{equation}
\label{Eq 8}
\begin{split}
       \rho_s \partial_t \textbf{\emph{v}}_s  + 
 \rho_s \textbf{\emph{v}}_s \cdot\boldsymbol{\nabla} \textbf{\emph{v}}_s= \boldsymbol{\nabla}  p_{s}^{1}  +\eta\boldsymbol{\nabla}^2\textbf{\emph{v}}_s  \\
 - E_{ac}\cos({2kx})\boldsymbol{\nabla}\hat{Z}
     + \sigma k\boldsymbol{\nabla} H,
     \end{split}
\end{equation}

where $p_{s1} = p_s+ {\phi}_{ac}$. It is important to note that assuming the amplitude of acoustic fields or $E_{ac}$ remains steady (i.e., unaffected by time or motion of the droplet) in Eq. \ref{Eq 8} provides a significantly simpler theoretical approach for studying the droplets compared to standard method of solving two-way coupled fast and slow time equations. This constant $E_{ac}$ assumption allows us to directly solve the slow time scale equations (Eq. \ref{Eq 8}, in conjunction with the continuity equation (Eq. \ref{Eq 4a}) and transport equation (Eq. \ref{Eq 4c}) without needing to solve the fast time scale equations. This approach significantly reduces computational time while still capturing all essential qualitative results, as shown in Fig. \ref{Fig 3} (a and b). Although the governing equation encompasses all aspects of droplet phenomena,  the non-dimensional acoustic Bond number obtained from the momentum equation $\left({\emph{f}}_{ac1}/{\emph{f}}_{\sigma}= E_{ac}\cos(2kx)(Z_2 - Z_1)/(\sigma k\right))$ completely fails to characterize droplet behaviours (including migration, deformation, and splitting). For example, the above predicts maximum acoustic force at node ($x=0$) and antinode ($x=\pi$), which is clearly not the case. The fundamental problem lies in the fact that $\textbf{\emph{f}}_{ac1}$ and $\textbf{\emph{f}}_{\sigma}$ do not entirely compete with each other. Instead, a considerable portion of both $\textbf{\emph{f}}_{ac1}$ and $\textbf{\emph{f}}_{\sigma}$ contributes solely to pressure. To address this, one can take the curl of the entire equation, which effectively eliminates the pressure contributions, resulting in
\begin{equation}
\label{Eq 9}
\begin{split}
      \boldsymbol{\nabla}\times (\rho_s \partial_t \textbf{\emph{v}}_s ) + \boldsymbol{\nabla}\times
 (\rho_s \textbf{\emph{v}}_s \cdot\boldsymbol{\nabla} \textbf{\emph{v}}_s)=\boldsymbol{\nabla} \times \left(\eta\boldsymbol{\nabla}^2\textbf{\emph{v}}_s \right) \\
 - \boldsymbol{\nabla}\times(E_{ac}\cos({2kx})\boldsymbol{\nabla}\hat{Z}
      \ ) +\boldsymbol{\nabla}\times( \sigma \kappa \boldsymbol{\nabla} H \ ).
      \end{split}
\end{equation}
Eventually, the acoustic and interfacial forces balance each other, bringing the system to equilibrium or rest ($\textbf{\emph{v}}_s=0$) as shown in Fig. (\ref{Fig 1} and \ref{Fig 3}), and the above equation simplifies to,
\begin{equation}
\label{Eq 10}
   - \boldsymbol{\nabla}\times(E_{ac}\cos({2kx})\boldsymbol{\nabla}\hat{Z}
      \ ) +\boldsymbol{\nabla}\times( \sigma \kappa \nabla H \ ) = 0.
\end{equation}
 The dimensionless form of the above equation results in ($x^\ast={x}/{L_{c}}$,\ $y^\ast={y}/{L_{c}}$, and $\kappa ^\ast={\kappa}{L_{c}}$),
\begin{subequations}
\label{Eq 11}
    \begin{equation} 
\label{Eq 11a}
 Bo\frac{\partial H}{\partial y^\ast}= 
 \left( \frac{\partial \kappa^\ast }{\partial x^\ast}\frac{\partial H}{\partial y^\ast}-\frac{\partial \kappa^\ast}{\partial y^\ast}\frac{\partial H}{\partial x^\ast}\right),
\end{equation}
\begin{equation} 
\label{Eq 11b}
Bo =\frac{2kE_{ac}\sin(2kx)\Delta \hat{Z}}{\sigma /L_{c}^2},
\end{equation}
\end{subequations}
 where $\Delta \hat{Z}=(Z_{2}-Z_{1})/Z_{avg}$. The acoustic Bond number ($Bo$) obtained from the vorticity equation completely captures the interplay between the acoustic and interfacial tension forces. Remarkably, it agrees with the acoustic Bond number governing the stability of immiscible fluids under acoustic fields \cite{Rajendran2023Jun}. Before demonstrating its accuracy in characterizing droplet deformation and splitting, it is crucial to recognize that acoustics body force $\textbf{\emph{f}}_{ac1}$ (Eq. \ref{Eq 7}) can be alternatively expressed similar to the acoustic force component present in the above $Bo$ in \ref{Eq 11b}
 ($\textbf{\emph{f}}_{ac1} =-E_{ac}\cos({2kx})\boldsymbol{\nabla}\hat{Z}=-\boldsymbol{\nabla} \left(\cos({2kx}) \right)-2kE_{ac}\sin({2kx})\hat{Z} \boldsymbol e_x$, where $\boldsymbol e_x$ is the unit normal vector along the $x$ direction). Substituting the above along with interfacial force $\left(\textbf{\emph{f}}_{\sigma}= \sigma \kappa \boldsymbol{\nabla} H = \boldsymbol{\nabla} \left (\sigma \kappa H \right)-\sigma H\boldsymbol{\nabla} \kappa \right)$ in  Eq. \ref{Eq 8} becomes, 
\begin{equation}
\label{Eq 12}
\begin{split}
    \boldsymbol{\nabla} \rho_s \partial_t \textbf{\emph{v}}_s  + 
 \rho_s \textbf{\emph{v}}_s \cdot\boldsymbol{\nabla} \textbf{\emph{v}}_s = \boldsymbol{\nabla} p_{s2}+\eta\boldsymbol{\nabla}^2\langle\textbf{\emph{v}}_s\rangle \\ -2kE_{ac}\sin({2kx})\hat{Z} \boldsymbol e_x 
    - \sigma H\boldsymbol{\nabla} \kappa,        
\end{split}
\end{equation}
where $p_{s2}=p_{s1}+\cos({2kx}) \hat{Z} +\sigma \kappa H$. Solving  Eq. \ref{Eq 12} (Fig. \ref{Fig 3}c), yields the same results as  Eqs. \ref{Eq 8} (Fig. \ref{Fig 3}b).  Thus instead of using $\textbf{\emph{f}}_{ac1}$, we can use the following equation as an acoustic body force,
\begin{equation} 
\label{Eq 13}
\textbf{\emph{f}}_{ac2}=-2kE_{ac}\sin({2kx})\hat{Z} \boldsymbol e_x.
\end{equation}
 The advantage of $\textbf{\emph{f}}_{ac2}$ (Eq. \ref{Eq 13}) over $\textbf{\emph{f}}_{ac1}$ (Eq. \ref{Eq 7}) is that the structure of $\textbf{\emph{f}}_{ac2}$ is consistent with the governing non-dimensional Bond number (Eq. \ref{Eq 11b}) derived from the vorticity equation as well as the structure of Gorkov's force Equation ($\boldsymbol{F}_{ac}$) for small particles, and it also facilitates better scaling. The velocity field $\textbf{\emph{v}}_s$ remains unchanged within the closed domain, regardless of whether Eq. \ref{Eq 7} or Eq. \ref{Eq 13} is used. The only difference is in the renormalization of the pressure field. 
 
 Eq. \ref{Eq 12} containing $\textbf{\emph{f}}_{ac2}$ is particularly noteworthy as it enables an intuitive and straightforward prediction/explanation of droplet behavior as follows. Initially, with the acoustic field turned off and the droplet being spherical or circular ($\kappa =$ constant), the interfacial force term ($\sigma H\boldsymbol{\nabla} \kappa $) is zero. This results in only Laplace pressure difference across the interface without any droplet/fluid motion. When an acoustic field is applied, acoustic force ($\textbf{\emph{f}}_{ac2}$) pushes the high-impedance fluid towards the node (low to antinode), as illustrated in Fig. \ref{Fig 1}. In this process $\kappa$ varies spatially, $\sigma H\boldsymbol{\nabla} \kappa $ becomes non-zero and counteracts $\textbf{\emph{f}}_{ac2}$. This deformation of the droplet continues till these forces balance each other. If the $\textbf{\emph{f}}_{ac2}$ is weak for the given $\sigma$ and $R$ ($Bo \ll 1$), the droplet migrates to node, like rigid particles as the curvature gradient is negligible (Fig. \ref{Fig 1}c). If $\textbf{\emph{f}}_{ac2}$ is strong for the given $\sigma$ and $R$ ($Bo \gtrsim 1$), then droplet deformation will be significant (Fig. \ref{Fig 1}a). This deformation can eventually lead to droplet splitting if multiple nodes are present within the droplet (Fig. \ref{Fig 1}b).


 
 
\begin{figure}[h!]
  \center
    \includegraphics[width=1\linewidth]{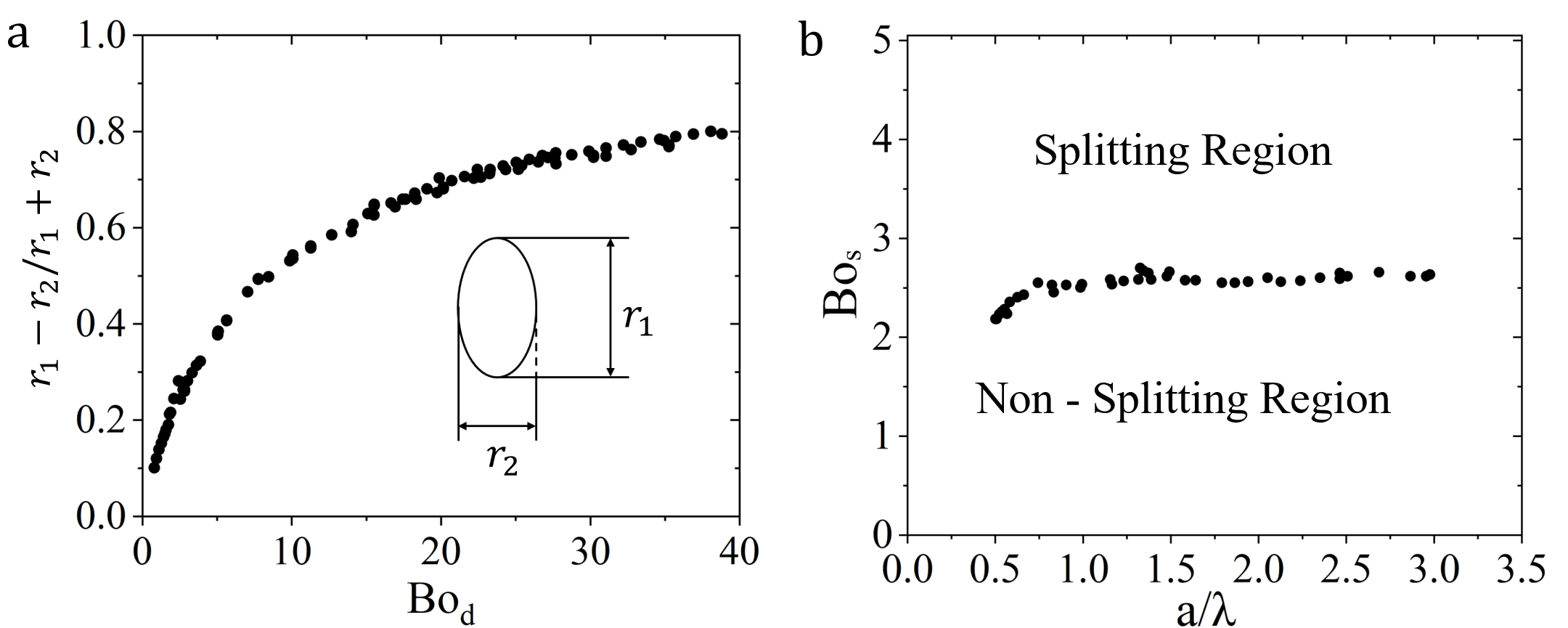} 
    \caption{a) Characterization of droplet deformation by $Bo_d$, with parameters varied in the range $\lambda$ (335 to 1340 \textmu m), $D$ (100 to 500 \textmu m),  and $E_{ac}$ (50 to 3000  J/m$^{3}$). b) Characterization of droplet Splitting by $Bo_s$. The critical $E_{ac}$ required to break the droplet is determined by varying the parameters $\lambda$ (223 to 1340 \textmu m), $D$ (150 to 1330 \textmu m), and $E_{ac}$ (400 to 2000 J/m$^{3}$). The data points separating the splitting and non-splitting region show the critical acoustic bond number required for splitting. Simulation data points are obtained from a simplified framework.} 
    
\label{Fig 4}
\end{figure}
Here, we proceed to show how the $Bo$ (Eq. \ref{Eq 11b}) characterizes droplet deformation and splitting by using the appropriate $L_c$. For droplet deformation when high impedance droplet is accommodated between two antinodes ($D<\lambda/2$), the relevant characteristic length becomes droplet radius $L_{c}=R$. Since as $R$ increases, interfacial forces becomes weaker. The force acting on the portion of the droplet present at node and antinode is zero, thus appropriate to take $\sin(2kx)\approx \sin(2k\bar{R})$, where $\bar{R}$ is the centroid of the droplet. The acoustic bond number ($Bo_d$) governing the droplet deformation becomes (Fig. \ref{Fig 4}a), 
\begin{equation} 
\label{Eq 14}
Bo_d =\frac{2kE_{ac}\sin(2k\bar{R})\Delta \hat{Z}}{\sigma /{R}^2}.
\end{equation}
For splitting, the relevant characteristic length $L_{c}$ is $\lambda/2$ as acoustic force tries to break/deform the bigger droplet into multiple smaller droplets of size $\lambda/2$ as shown in Fig. \ref{Fig 3}. Since the bigger droplet covers multiple nodes and antinodes, the droplet's position and initial radius will not play any major role in splitting, $\sin(2kx)\approx 1$, the bond number governing the splitting becomes (as shown in Fig. \ref{Fig 4}b),  
\begin{equation} 
\label{Eq 15}
Bo_s =\frac{2kE_{ac}\Delta \hat{Z}}{\sigma /(\frac{\lambda}{2})^2}.
\end{equation}
After splitting, $Bo_d$ further characterizes the deformation of the daughter droplets. 

Discussion: This work can serve as a fundamental tool for studying droplets in acoustic fields. The presented theoretical framework can be employed to study a range of acoustic phenomena, including the acoustic levitation of droplets (such as migration, deformation, and breakup), acoustic fountains, acoustic streaming within droplets, and droplet coalescence, among other phenomena. 


Acknowledgement:  This work is supported by  the Department of Science \& Technology - Fund for Improvement of Science \& Technology Infrastructure (DST-FIST) via Grant No: SR/FST/ET-I/2021/815.

$^{\dagger}$  These authors have contributed equally to this work. 
\bibliography{references}

\end{document}